\documentclass{article}

    \PassOptionsToPackage{numbers, compress}{natbib}

 \usepackage[preprint]{tackling_climate_workshop_style}

\usepackage[utf8]{inputenc} 
\usepackage[T1]{fontenc}    
\usepackage{hyperref}       
\usepackage{url}            
\usepackage{booktabs}       
\usepackage{amsfonts}       
\usepackage{nicefrac}       
\usepackage{microtype}      

\usepackage{comment}

\title{AI, Climate, and Transparency:\\Operationalizing and Improving the AI Act}

\renewcommand{\thefootnote}{\fnsymbol{footnote}}
\author{%
  Nicolas Alder\footnotemark[2]  \\
  Hasso Plattner Institute\\
  Potsdam, Germany \\
  \texttt{nicolas.alder@hpi.de} \\
   \And
  Kai Ebert\footnotemark[2] \\
  European University Viadrina\\
  Frankfurt (Oder), Germany \\
  \texttt{kebert@europa-uni.de} \\
   \AND
  Ralf Herbrich \\
  Hasso Plattner Institute\\
  Potsdam, Germany \\
  \texttt{ralf.herbrich@hpi.de} \\
   \And
  Philipp Hacker \\
  European University Viadrina\\
  Frankfurt (Oder), Germany \\
  \texttt{hacker@europa-uni.de} \\
}

\begin{document}

\footnotetext[2]{Equal contribution.}

\maketitle
\renewcommand{\thefootnote}{\arabic{footnote}} 
\setcounter{footnote}{0}  

\begin{abstract}
  This paper critically examines the AI Act's provisions on climate-related transparency, highlighting significant gaps and challenges in its implementation. We identify key shortcomings, including the exclusion of energy consumption during AI inference, the lack of coverage for indirect greenhouse gas emissions from AI applications, and the lack of standard reporting methodology. The paper proposes a novel interpretation to bring inference-related energy use back within the Act’s scope and advocates for public access to climate-related disclosures to foster market accountability and public scrutiny. Cumulative server level energy reporting is recommended as the most suitable method. We also suggests broader policy changes, including sustainability risk assessments and renewable energy targets, to better address AI's environmental impact.
\end{abstract}

\section{Introduction}
\label{sec:introduction}
The climate implications of artificial intelligence (AI), including energy and water consumption, are increasingly subjected to public scrutiny and academic research \cite{luccioni2023estimating, li2023making, dodge2022measuring, kaack2022aligning, strubell2020energy, hacker2024sustainable}. While energy efficiency targets for data centers are under discussion,\footnote{Further EU legislation can be expected follwing the Commission's report by 15 May 2025 under Art. 12(5) of the Energy Efficiency Directive (EU) 2023/1791.} there is concern that their energy consumption could surpass the available supply of renewable energy. Major companies like Google have reported that increased energy demand related to AI endangers their carbon zero strategies\footnote{\href{https://www.gstatic.com/gumdrop/sustainability/google-2024-environmental-report.pdf}{Google 2024: Environmental Report}}.

As in many collective action problems, regulation may play a major part in mitigating the negative impact of AI on climate while fostering socially beneficial use cases. Globally, several initiatives are underway to establish legal frameworks for AI. The most prominent example is probably the EU AI Act, which has just entered into force at the beginning of August 2024. The Act also includes significant sections concerning climate impacts, primarily reporting obligations. Thus, one might hope, the climate effects of AI could become a relevant market parameter; have reputational repercussions; and enable public scrutiny, by analysts and NGOs.

Against this background, this paper analyzes the Act's transparency provisions from both a legal and a technical perspective, and makes three core contributions. First, it shows that the Act falls short in several critical areas. Second, we argue that even within the AI Act's current scope, operationalizing its mandates presents significant challenges. Third, we make a range of policy proposals that seek to address these challenges. This necessity extends beyond the AI Act to broader policy changes, including the Global Digital Compact currently under discussion at the United Nations.
 
\section{Climate Transparency and the AI Act: Gaps and Interpretation Challenges}
\label{sec:climate_and_the_ai_act}

Any change for the better starts with information about what is wrong. However, at the moment, it is often unclear what the exact impact of the development and usage of an AI model is concerning energy and water consumption. The AI Act seeks provide a remedy by forcing certain AI providers to make climate-related disclosures. However, the patchwork of provisions includes seven significant ambiguities and loopholes.

First, for high-risk AI systems, providers are required under Art. 11(1) to document the computational resources used in development, training, testing, and validation, as per Annex IV(2). However, there is no explicit requirement to disclose energy consumption, limiting the comparability of, and transparency on, the environmental impact of these high-risk systems to estimates based on the documented computational resources.

Second, the AI Act imposes transparency obligations on providers of general-purpose AI (GPAI) models, particularly concerning energy consumption. Under Art. 53(1)(a), providers must maintain up-to-date technical documentation that includes information specified in Annex XI, which requires known or estimated energy consumption of the model, with estimates potentially based on computational resources. However, this requirement focuses on the model's development phase, excluding the inference phase, which is a significant oversight given the potentially much greater cumulative energy consumption during inference \cite{luccioni2024power, wu2022sustainable}.
To address this gap, a novel interpretation can be considered. Art. 53(1)(a) and (b), in conjunction with Annex XI and Annex XII, require providers to include in the documentation for downstream AI system providers and authorities information on the technical means needed to integrate the GPAI model into AI systems. Although energy consumption is not explicitly mentioned, these provisions should, arguably, be interpreted to include information on hardware requirements, allowing downstream providers to estimate the energy consumption for inference. This novel interpretation would indirectly ensure transparency regarding inference energy use.

A third issue arises with open-source (OS) GPAI models, which are generally exempt from transparency obligations unless they pose a systemic risk (Art. 53(2)). Recital 102 emphasizes transparency for OS models but does not include energy consumption in the information that must be disclosed. Rather, the focus is on parameters, model architecture, and usage information, leaving a gap in transparency regarding the energy impact of these models. 

Regarding, fourth, fine-tuning, Recital 97 seems to imply that an entity engaging in any, even minuscule, fine-tuning of a GPAI model automatically becomes the provider of a new model, with all corresponding duties. For minor changes, this seems excessive, even though Recital 109 suggests that reporting obligations are limited to that fine-tuning. However, Art. 25(1)(b) holds that, for high-risk AI systems (e.g., in recruitment), only a substantial modification bestows provider status upon the modifying entity. This rule could be analogized for fine-tuning, such that only substantial model modifications via fine-tuning lead to provider status, protecting smaller entities.

Fifth, the AI Act overlooks the greenhouse gas (GHG) effects of AI applications, such as those used in oil and gas exploration \cite{kaack2022aligning}. This omission leaves a significant gap, as these applications can substantially contribute to climate change, yet their environmental impact remains unreported. 

Sixth, while the Act requires energy consumption to be documented, this information is only available to authorities, not downstream providers (unless our suggested interpretation is adopted), and not to the general public. Without broader access to this data, transparency and accountability are significantly curtailed, hindering market effects based on climate reporting, independent research and verification, and public scrutiny by analysts and NGOs.

Finally, the Act also fails to address the use of toxic materials and water consumption, a critical factor in data center operations. While most data centers in the EU must report their water usage under the Energy Efficiency Directive, the AI Act lacks a specific attribution to AI, as stipulated for energy consumption, and computing outside the EU is not covered. Given the significant water usage for cooling in data centers, this omission leaves a major aspect of AI's environmental impact unreported.


\section{Operationalizing the Requirements: Implementation Challenges}

As the previous section showed, under the current version of the AI Act, GPAI providers must log the energy consumption used for training GPAI models. To operationalize this provision, it is crucial to clarify how energy consumption should be measured or estimated. We discuss three methods: measurement at the data center level; at the cumulative server level; and at the individual graphic-processing unit (GPU) level.

Energy efficiency in data centers is measured by the Power Usage Efficiency (PUE) metric. It denotes the ratio of total energy used by the data center to the energy consumed by its computational hardware. A lower PUE indicates higher energy efficiency, with a global average PUE of 1.58 recorded in 2023\footnote{\href{https://www.statista.com/statistics/1229367/data-center-average-annual-pue-worldwide/}{Statista 2023: What is the average annual power usage effectiveness (PUE) for your largest data center?}}.
When measuring energy consumption at the data center level, the advantage lies in capturing the total power usage, including both direct computing energy and overhead like cooling. This provides a comprehensive overview and encourages efficient data center selection. However, it can obscure the energy impacts of specific model architecture or software inefficiencies, as these are influenced by the data center's overall efficiency. Estimating with the PUE ratio is practical but may lack precision for specific model-level insights.

At the cumulative server level, i.e., for all utilized servers within one data center, energy measurement with power distribution units is highly accurate, closely reflecting model size, data volume, and software efficiency. This method is recognized in the industry and can provide detailed insights into energy consumption. However, not all data centers currently track power demand at this level, and implementing such systems can be time-consuming\footnote{\href{https://www.thegreengrid.org/en/resources/library-and-tools/572-IT-\%26-Power-Efficiency-Survey-Results-\%28for-EUC\%29}{The Green Grid 2023: IT \& Power Efficiency Survey Results}}. While cloud providers like AWS and Azure may have these capabilities, widespread reporting standards are lacking, potentially disadvantaging smaller companies.

Finally, measuring energy usage at the GPU level within a server is straightforward with on-chip sensors for components like NVIDIA GPUs, which offer user-friendly monitoring. However, this approach significantly underestimates total energy consumption as it only accounts for a single component, missing the broader picture of server-wide energy use. Therefore, it is not recommended for comprehensive energy tracking.

\section{Discussion and Policy Proposals}

The AI Act is a first step toward mandatory AI related climate reporting, but is riddled with loopholes and vague formulations. To remedy this, we make six key policy proposals. Such mechanisms should not only be included in the evaluation report due in August 2028 (Art. 111(6)), but in any interpretive guidelines by the AI Office and other agencies, reviews and potential textual revisions beforehand. 

The primary weakness of the AI Act is the exclusion of inferences from explicit and mandatory energy consumption reporting. While we offer a solution for interpretation, it is unclear whether courts, agencies and companies will follow this route. This significantly hampers the assessment of future AI energy usage, related carbon emissions, and effects on (renewable) energy infrastructure. Hence, future guidance from the AI Office, and delegated acts by the Commission (Art. 53(5) and (6)), should explicitly include inference as a reporting category, both in Annex XI (for the AI office) and XII (for downstream actors).


Another major challenge is the failure to include indirect emissions by AI applications (e.g., for oil and gas exploration) and water consumption within the reporting obligations. This should be remedied at the provider (water) and the deployer level (applications). 

Third, the consequences of minor fine-tuning operations on GPAI remain unclear. It would be beneficial to tie the energy reporting requirement to the mechanism of training (fitting model weights) and incorporate a minimum computational cost threshold, as this would encompass energy-intensive training and fine-tuning for reasonably sized workloads.

The open source exemption, fourth, should be revoked. There is no convincing reason to abstain from climate reporting only because other parts of the model are made public and transparent.


Fifth, energy consumption measurements ought to be conducted at the cumulative server level and reported accordingly. This reflects the total computation-related power usage. Furthermore, the PUE factor of each data center, as measured and reported under the Energy Efficiency Directive (EU) 2023/1791 and Delegated Regulation (EU) 2024/1364, provides information for a relevant estimate of the overall energy consumption. By reporting these numbers separately, we can differentiate between the power usage specific to the model (server-level computation) and the efficiency of the data center, thus reflecting the realistic overall energy investment. Estimations for server-level power consumption should utilize peak utilization values from the hardware manufacturer (e.g., NVIDIA). When actual measurements are available, they must be prioritized over estimations. The ultimate aim is to secure as precise power consumption data as possible, allowing for flexibility for model providers with limited access to data infrastructure (such as for finetuning with substantial modification), while also ensuring that estimations are not misused to avoid accurate measurement reporting. These considerations should inform both the technical standards drafted under Art. 40 AI Act and the possible implementation of the Global Digital Compact at the international level.

Finally, sixth, all climate-related disclosures must urgently be made available to the general public, not only to authorities and, potentially, downstream actors. Trade secrets and intellectual property do not stand in the way if only aggregate numbers at the cumulative server level are reported. Only in this way, market pressure can build up, reputational effects set in, and public scrutiny via analysts, academics, and NGOs unfold its incentivizing force.

\begin{table}[t]
  \caption{Shortcomings in the AI Act Concerning Climate Reporting, and Policy Proposals}
  \label{table:climate_reporting_gaps}
  \centering
  \begin{tabular}{p{3.5cm}p{9cm}l}
    \toprule
    \cmidrule(r){1-2}
    Shortcomings & Policy Proposals \\
    \midrule
    1. Inference Energy Consumption Exclusion & Explicitly include inference in energy reporting obligations in Annexes XI and XII. \\
    2. Indirect Emissions and Water Consumption & Extend reporting obligations to include water consumption and indirect GHG emissions from AI applications.\\
    3. Fine-Tuning Uncertainty & Clarify uncertainty of reporting obligations by tying them to computational cost and training mechanisms. \\
    4. Open-Source Models & Revoke the exemption to ensure comprehensive climate reporting. \\
    5. Lack of Standard Reporting Methodology & Measure energy consumption at the cumulative server level, with separate PUE.  \\
    6. Lack of Public Access to Energy Data & Make all climate-related disclosures publicly available to foster transparency and market accountability. \\
    \bottomrule
  \end{tabular}
\end{table}

\section{Conclusion}

This paper tackles some of the complexities at the intersection of AI, climate and regulation. The AI Act does contain significant climate reporting obligations. By drawing on technical and legal research, we show that they contain too many loopholes, and are difficult to operationalize. Perhaps most importantly, even though recent research has shown inference to be a major driver of AI-related GHG emissions, this key area is omitted from the AI Act. A novel interpretation of the Act’s reporting obligations might bring inference back within its scope. Furthermore, none of the climate disclosures are initially open to the public. We suggest changing this urgently to kickstart market pressure, induce reputational effects, and enable crucial public scrutiny, e.g. by academics and NGOs.

However, climate reporting can only be a first step in addressing the massive and fast-rising environmental impact of AI models and systems. It must be complemented by substantive obligations, including sustainability risk assessment and management, renewable energy targets for data centers, and potentially even (tradable) caps on the energy and water consumption of data centers and similar major consumption drivers in the AI value chain \cite{hacker2024sustainable}.

\bibliographystyle{plainnat}
\bibliography{literature}
\end{document}